\documentclass[12pt]{article}
\usepackage{geometry}
\usepackage{amsmath,amssymb}
\usepackage[nosort]{cite}
\usepackage{float}
\usepackage{latexsym}
\usepackage{graphicx}
\usepackage{color}
\usepackage{subfig}
\usepackage{color}

\newcommand{\beq}{\begin{equation}}
\newcommand{\be}{\begin{equation}}
\newcommand{\ee}{\end{equation}}
\newcommand{\bea}{\begin{eqnarray}}
\newcommand{\eea}{\end{eqnarray}}

\newcommand{\pa}{\partial}

\newcommand{\nn}{\nonumber}

\newcommand{\mc}[1]{{\mathcal #1}}

\newcommand{\Dh}{\hat{D}}
\newcommand{\Qh}{\hat{Q}}
\newcommand{\mf}{\mathcal F}
\newcommand{\ma}{\mathcal A}

\begin{document}

\begin{titlepage}
\hbox to \hsize{\hspace*{0 cm}\hbox{\tt }\hss
    \hbox{\small{\tt }}}

\vspace{1 cm}

\centerline{\bf \Large Thermal behavior of charged dilatonic black branes in AdS}

\vspace{.6cm}

\centerline{\bf \Large and}

\vspace{.6cm}

\centerline{\bf \Large  UV completions of Lifshitz-like geometries}

\vspace{1 cm}
 \centerline{\large  $^\dagger\!\!$ Gaetano Bertoldi, $^\dagger\!\!$ Benjamin A. Burrington and $^\dagger\!\!$ Amanda W. Peet}

\vspace{0.5cm}
\centerline{\it ${}^\dagger$Department of Physics,}
\centerline{\it University of Toronto,}
\centerline{\it Toronto, Ontario, Canada M5S 1A7. }

\vspace{0.3 cm}

\begin{abstract}
Several classes of gravitational backgrounds in $3+1$ dimensions have been proposed as holographic duals to Lifshitz-like theories describing critical phenomena in $2+1$ dimensions with critical exponent $z\geq 1$.  We numerically explore one such model, characterized by a temperature $T$ and chemical potential $\mu$, and find how to embed these solutions into AdS for a range of values of $z$.  We find no phase transition going from the $T\ll\mu$ to the $T\gg \mu$ regimes, and find that the solutions smoothly interpolate between the Lifshitz-like behavior and the relativistic AdS-like behavior.  Finally, we exploit some conserved quantities to find a relationship between the energy density $\mc E$, entropy density $s$, and number density $n$, $\mc E=\frac{2}{3} \left(Ts+\mu n\right)$.  We show that this result is expected from general scaling arguments, and generalizes to $\mc E= \frac{d}{d+1}\left(Ts+\mu n\right)$ for a theory dual to AdS$_{d+2}$ (Poincar\'e patch) asymptotics with a local $U(1)$ gauge invariance.
\end{abstract}
\end{titlepage}

\section{Introduction}

Since the Maldacena conjecture \cite{Maldacena:1997re}, AdS/CFT has become an important tool used to study non perturbative aspects of field theories (for a review see \cite{Aharony:1999ti}).  Traditionally, this has been used to study $3+1$ superconformal field theories in a particle physics context \cite{Oz:1998hr}, however considerably less symmetric theories can be studied using holographic techniques \cite{smallsym}.  It has also been fruitful to study toy models \cite{toys} that imitate certain features of string theory backgrounds, and seem to contain much of the physics\footnote{Of course one always prefers string theory models where one can directly state what the weak coupling degrees of freedom are.}.

From the $3+1$ dimensional cases, it has been learned that finite temperature corresponds to the presence of a horizon in the gravitational dual.  Further, $U(1)$ gauge symmetries in the bulk correspond to conserved number operators in the dual field theory.  Therefore, to study a field theory at finite temperature and chemical potential in a holographic setup, one studies charged objects with horizons in the bulk: charged black holes \cite{bhphases}.

Holographic techniques have recently been applied to lower dimensional non-relativistic systems as well \cite{Son:2008ye,Balasubramanian:2008dm,condensedmatter} (for reviews see \cite{Hartnoll:2009sz}).  In particular, much effort has gone into describing quantum critical behavior of these theories.  Quantum critical systems exhibit a scaling symmetry
\be
t\rightarrow \lambda^{z}t, \quad x_i\rightarrow \lambda x_i
\ee
similar to the scaling invariance of pure AdS ($z=1$) in the Poincar\'e patch.  From a holographic standpoint, this suggests the form of the spacetime metric
\be
ds^2=L^2\left(r^{2z} dt^2+ r^2 dx^i dx^j \delta_{i j}+\frac{dr^2}{r^2}\right), \label{scalemetric}
\ee
where the above scaling is realized as an isometry of the metric along with $r\rightarrow \lambda^{-1} r$ \footnote{Earlier studies of these metrics in a ``brane world'' scenario appear in \cite{Koroteev:2007yp}.  It is also interesting to note that this metric, as well as several other ``non-relativistic'' metrics, are coset spaces \cite{SchaferNameki:2009xr}.}.  The above metric has no symmetry that mixes time and space, although there are also models that contain a Galilean (or other non-relativistic) symmetry \cite{Balasubramanian:2008dm,Yamada:2008if,Bobev:2009mw}.  However, here we will be concerned with models that admit the metric (\ref{scalemetric}) as a solution.

Often, a good place to begin studying any system is to write down a toy or ``phenomenological'' model \cite{Son:2008ye,Balasubramanian:2008dm} to study generic properties.  One may then consider possible embeddings into a more fundamental theory \cite{Yamada:2008if,Azeyanagi:2009pr,Hartnoll:2009ns,Blaback:2010pp}, such as a string theory, where more information is known about the weakly coupled physics.  Two toy models that admit the metric (\ref{scalemetric}) have Lagrangians given by \cite{Kachru:2008yh}
\be
S'=\frac{1}{16\pi G_4}\int d^4x\sqrt{-g}\left(R-2\Lambda-\frac14 \mf^2-\frac{c^2}{2}\ma^2\right)
\ee
(we call this model $S'$ for short) and \cite{Taylor:2008tg}
\be
S=\frac{1}{16 \pi G_4}\int d^4x \sqrt{-g}\left(R-2\Lambda-2(\nabla \phi)^2-e^{2\alpha \phi} {\mathcal F}^2\right).
\ee
(we call this model $S$ for short) where in either case $\mf=d \ma$ is a two form field strength.  There are also examples of actions with $R^2$ corrections that admit solutions of the form (\ref{scalemetric}) \cite{Cai:2009ac}.

There are several differences between the above models.  First, the Lifshitz solution to model $S'$ is truly invariant under the Lifshitz rescaling symmetry, while the solution with metric (\ref{scalemetric}) of model $S$ has a logarithmically running dilaton, and so we call this a ``Lifshitz-like'' geometry.  Note also that $S$ has a $U(1)$ gauge symmetry while $S'$ does not, and so the field theory dual to $S$ has a conserved particle number $N$ with chemical potential $\mu$.  Also of note is that model $S$ admits an exact black brane solution \cite{Taylor:2008tg} that asymptotes to the metric (\ref{scalemetric}), which can also be generalized to higher dimensions \cite{Chen:2010kn}.  Finding black brane solutions for action $S'$ has proven more difficult, and one often needs to resort to numeric methods \cite{Danielsson:2008gi,Mann:2009yx,Bertoldi:2009vn}, but not always \cite{Bertoldi:2009dt} (an analogous analytic statement for an $R^2$ extension may be found in \cite{Dehghani:2010gn}, and one should also see \cite{Balasubramanian:2009rx} where a certain extension to this model admits an analytic black hole).  For extensions and variations to the basic model $S'$, see \cite{Dehghani:2010gn,extensions}, and for the holographically renormalized action, see \cite{Ross:2009ar}.

In our current work we study the action $S$, also studied in \cite{Taylor:2008tg,Pang:2009ad,Goldstein:2009cv} (for variations and extensions, see \cite{Chen:2010kn,Aprile:2010yb,Charmousis:2010zz,Liu:2010ka,Fadafan:2009an,Sin:2009wi,Cadoni:2009xm}).   In the Lifshitz-like solutions, the gauge field, and in fact the term $e^{2\alpha \phi} \mf^2\sqrt{-g}$ in the Lagrangian, diverges.  One immediately suspects that one must add a counter term on the boundary of the form $*\mf \wedge \ma$, and one would expect this to change the ensemble from a fixed $U(1)$ potential to a fixed $U(1)$ charge (again, fixed $N$ in the boundary theory).  One can also consider a fixed charge ensemble to simply be a limit where the chemical potential $\mu$ is large compared to other scales in the theory.  Indeed, this is realized holographically by a geometry that interpolates between the Lifshitz-like solution and AdS; the large $\mu$ limit corresponds to a limit that leaves only the Lifshitz-like geometry.

We now consider such a solution.  We take that the full geometry is the Lifshitz-like geometry in the $r\rightarrow 0$ limit, and is AdS for $r \rightarrow \infty$.  This type of solution was already numerically studied in \cite{Goldstein:2009cv},  and is essentially a relativistic (actually conformally invariant) UV completion of the theory.  In the AdS asymptotics $\phi$ goes to a constant and $\ma$ goes to a constant, and so the growth in both $\phi$ and $\ma$ is cut off in this background.  The size of $e^{\alpha \phi} \ma_t$ is essentially cut off by the scale where the theory becomes like AdS.  Conversely, the scale in the geometry is determined by $e^{\alpha \phi} \ma_t$, and we regard this as the chemical potential $\mu_{\rm geom}=\mu L^2$ (where $L$ is the radius of AdS).  Therefore, for such a geometry, we expect to have two regimes for energy scales: $E \ll \mu$ and $E \gg \mu$.  For the regime $E \ll \mu$, the quantity $\mu$ provides a cutoff for new physics, i.e. it is the scale at which one can excite quanta associated with the conserved particle number, and can be considered the scale leading to the RG flow of $\phi$ in the Lifshitz-like background.  Further we can see that if we perform the rescaling $(t,x_i,r)\rightarrow (\lambda^{-1} t, \lambda^{-1} x_i, \lambda^{1} r)$ for large $\lambda$, we are zooming in on the Lifshitz-like part, and this is scaling $\mu\rightarrow \infty $.  Combined with a time rescaling, this limit will eventually leave only the Lifshitz-like geometry.

There is of course another natural energy scale that one can introduce: the temperature $T$.  From the above, we expect there to be two regimes $T \ll \mu$ and $T \gg \mu$, and we would like to know how the theory interpolates between these two known asymptotic solutions.  We were initially motivated to find a possible phase transition when going from one regime to the other, as $\mu$ offers a new scale.  Black hole phase transitions are typically first order; there are two solutions for a given set of boundary conditions, but each solution has different energy densities.  Hence, during the phase transition, one expects a non homogeneous phase (when considering a fixed energy ensemble). However, we find no such phase transition when studying the system numerically.  A possible explanation is that we are studying the vacua of a $2+1$ dimensional theory.  As the vacua are time independent, the space of vacua for the theory are defined by a $2$ dimensional field theory.  Such a field theory cannot have a phase transition that breaks the global translational and rotational symmetries (the inhomogeneous phase mentioned above) due to the Coleman$-$Mermin$-$Wagner (CMW) theorem (for a recent holographic study, see \cite{Anninos:2010sq}).  Therefore, one might expect no phase transition to occur (this leaves open the possibility that for greater values of the spatial dimension $d$ the theory might have different thermodynamic behavior).  In all our results, we find a smooth, monotonic change in behavior from Lifshitz-like behavior to AdS-like behavior, similar to the ``extremal'' $r_h\rightarrow 0$ case studied numerically in \cite{Goldstein:2009cv}.

We also explore the consequences of having certain conserved quantities along the flow, much in the same spirit as our earlier work \cite{Bertoldi:2009vn,Bertoldi:2009dt}.  Using these conserved quantities, we are able to show that the energy density satisfies
\be
\mc E=\frac23 \left(Ts + \mu n\right) \label{thermrel1}
\ee
where $s$ is the entropy density, and $n$ is the number density in the dual theory.  We argue that this relation is actually expected in AdS.  Essentially, the above relation along with the first law of thermodynamics allows one to write a pair of first order linear PDEs for $s(T,\mu)$ and $n(T,\mu)$.  The solutions to these differential equations have a scaling symmetry that reproduces the expected scaling dependence of AdS.  In fact, the scaling dependence of AdS is shown to be equivalent to relation (\ref{thermrel1}).  This argument generalize to
\be
\mc E=\frac{d}{d+1} \left(Ts + \mu n\right) \label{genthermorel}
\ee
in $d$ spatial dimensions (of the field theory), i.e. for a theory with AdS$_{d+2}$ asymptotics and a local $U(1)$ gauge theory in the bulk.  We believe these arguments to be general enough to apply to any theory with only two scales $\mu$ and $T$ that has a conformally invariant UV fixed point.

We present this work as follows:  in section 2 we perform a reduction of the action $S$ along a certain Ansatz and reproduce the equations of motion from this action.  We find that the system reduces to 4 first order ODEs after introducing certain first integrals (conserved quantities), and further comment on the normalization of one such quantity, $Q$.  We then perform a perturbative analysis of solutions near a regular horizon, around the AdS asymptotics, and around the Lifshitz-like solution, and display the exact AdS and Lifshitz black branes.  We then discuss the setup for the numeric integration, and comment on how to read the chemical potential from the background.  Finally, in section 3, we present the results of our numeric analysis, and the analysis that leads to equation (\ref{genthermorel}).

\section{Analysis of the model}

\subsection{Reduction}

We wish to consider dilatonic black brane solutions to the equations of motion following from the action
\be
S=\frac{1}{16 \pi G_4}\int d^4x \sqrt{-g}\left(R-2\Lambda-2(\nabla \phi)^2-e^{2\alpha \phi} {\mathcal F}^2\right). \label{baseact}
\ee
We reduce the above action on the following Ansatz
\bea
ds^2=-e^{2A(r)}dt^2+e^{2B(r)}\left((dx_1)^2+(dx_2)^2\right)+e^{2C(r)}dr^2 \nn \\
\phi=\phi(r), \qquad {\mathcal A}=e^{G(r)}dt
\eea
where ${\mathcal F}=d{\mathcal A}$.  We will get only ordinary differential equations in what follows, and so we define $\pa\equiv \frac{\pa}{\pa r}$.  We may reduce the action to a one dimensional action \footnote{we keep track of normalization for later use}
\be
S=\frac{1}{16 \pi G_4}\int 2 dt dx_1 dx_2 \int dr L_{1D}
\ee
 with lagrangian
\bea
&&L_{1D}=2e^{A+2B-C}\pa A \pa B+e^{A+2B-C}(\pa B)^2+e^{-A+2B-C+2G+2\alpha \phi}(\pa G)^2 \nn\\
&&\qquad \qquad \qquad \qquad \qquad \qquad \qquad \qquad\qquad -e^{A+2B-C}(\pa \phi)^2-e^{A+2B+C}\Lambda. \label{redlag}
\eea
It can be verified that all equations of motion associated with the action (\ref{baseact}) are reproduced by (\ref{redlag}) as long as one uses the equation of motion for $C$.  Here, $C$ acts as a Lagrange multiplier, imposing the ``zero energy'' condition.  Further, we note that $C$ allows for generic $r$ diffeomorphisms.  We will refer to changing $r$ coordinate as ``coordinate gauge'' transformations, to differentiate from the $U(1)$ gauge transformations associated with ${\mathcal A}$.

There are many conserved quantities associated with the above action.  First, there is the conserved quantity associated with the shift symmetry $(A,B,C,\phi,G)\rightarrow (A+2\delta_1,B-\delta_1,C,\phi,G+2\delta_1)$. This can be understood as a rescaling of the time coordinate, and the $x_i$ coordinates that leaves $dt dx_1 dx_2$ invariant, and this then descends to the 1D action as Noether symmetry.  Further, we note that there is the conserved quantity associated with $e^G\rightarrow e^G+{\rm const}$ which is exactly the gauge symmetry associated with ${\mathcal A}$.  Further, we note that $(A,B,C,\phi,G)\rightarrow (A,B,C,\phi+\delta_2,G-\alpha\delta_2)$ is also a symmetry.  One can view this as saying that the dilaton here is exactly a space dependent gauge coupling, and we may absorb a normalization of the gauge coupling into the definition of ${\mathcal A}$.  In addition, we have the Hamiltonian constraint as always.  Now we count $4$ dynamical fields, and $4$ conserved quantities!  This implies that the system is completely equivalent to a set of first order differential equations
\bea
&&2e^{A+2B-C}\pa A \pa B+e^{A+2B-C}(\pa B)^2+e^{-A+2B-C+2G+2\alpha \phi}(\pa G)^2 \nn\\
&&\qquad \qquad \qquad \qquad \qquad \qquad-e^{A+2B-C}(\pa \phi)^2+e^{A+2B+C}\Lambda=0 \\
&&e^{A+2B-C}\pa A-e^{A+2B-C}\pa B-2e^{-A+2B-C+2G+2\alpha\phi}\pa G= {\mathcal{D}}_0 \\
&&e^{A+2B-C}\pa \phi+\alpha e^{-A+2B-C+2G+2\alpha\phi}\pa G={\mathcal{P}}_0 \\
&&e^{-A+2B-C+G+2\alpha\phi}\pa G= Q.
\eea
Note, we still have not fixed a particular coordinate or $U(1)$ gauge.  While there are many conserved quantities above, one can see that they do not Poisson commute, and so the above is not an integrable system.

Now, let us simplify things a bit, and replace $\pa G$ by $Q$ using the last of the above relationships.  We further take a different linear combination of the above equations, and find
\bea
&&2e^{A+2B-C}\pa A \pa B+e^{A+2B-C}(\pa B)^2+e^{A-2B+C-2\alpha \phi}Q^2 \nn\\
&&\qquad \qquad \qquad \qquad \qquad \qquad-e^{A+2B-C}(\pa \phi)^2+e^{A+2B+C}\Lambda=0 \\
&&2e^{A+2B-C}\left(2\pa \phi+\alpha(\pa A-\pa B)\right)= D_0 \\
&&e^{A+2B-C}\pa \phi+\alpha Q e^{G}= P_0\\
&&e^{-A+2B-C+G+2\alpha\phi}\pa G= Q
\eea
where we have redefined the integration constants $4{\mathcal{P}}_0+2\alpha{\mathcal{D}}_0\equiv D_0$, ${\mathcal{P}}_0\equiv P_0$ for convenience.  We note that $P_0$ is the only $U(1)$ gauge dependent quantity, i.e. it transforms under the map $e^G\rightarrow e^G+{\rm const}$, while the other equations do not.  In the above, we have used the notation $Q$ for a non normalized charge density.

Notice that the total contribution from the ${\mathcal F}^2$ term is
\be
S_{{\mathcal F}^2}=\frac{1}{16 \pi G_4}\int 2 dt dx_1 dx_2 \int dr \pa(e^{G}) Q.
\ee
When we compactify along an imaginary time coordinate, the integral over $t$ simply becomes the inverse (unitless) temperature, and we restrict to a finite patch in $x_1$ and $x_2$, so that this part of the action reads
\be
S_{{\mathcal F}^2}=\frac{1}{16 \pi G_4}2 \frac{\Delta x_1 \Delta x_2}{\hat{T}}Q [e^{G}]|^{r_{\infty}}_{r_h}
\ee
when evaluated in a black hole background where $r_h$ is the radius of the horizon.  We note that this contribution is therefore
\be
S_{{\mathcal F}^2}=\frac{1}{16 \pi G_4}2Q \frac{\Delta x_1 \Delta x_2}{\hat{T}}[e^{G}]|^{r_{\infty}}_{r_h}=\frac{1}{16 \pi G_4}2Q\frac{\Delta x_1 \Delta x_2}{\hat{T}}\mu_{\rm geom},
\ee
where $\mu_{\rm geom}$ is the potential difference between the boundary and the horizon.  Because we are using unitless $t,x_i$, we will have to restore units with $L$ in this expression via $T=\frac{\hat{T}}{L}$, $V_2=L^2 \Delta x_1 \Delta x_2$, to find
\be
S_{{\mathcal F}^2}=\frac{1}{16 \pi G_4}2Q \frac{V_2}{L^2}\frac{1}{LT}\mu_{\rm geom}.
\ee
This term in the action should give rise to a term $2S_{{\mathcal F}^2}=\frac{1}{T}\mu_{\rm geom} q_{\rm geom} V_2$ (see the general discussion in \cite{Batrachenko:2004fd}) where $q_{\rm geom}$ is a properly normalized charge density.  Therefore, we identify
\be
Q=\frac{q_{\rm geom} 16 \pi G_4 L^3}{4}
\ee
with $q_{\rm geom}$ properly normalized (we have chosen the charge convention that when $\mu_{\rm geom}$ is positive, so is $q_{\rm geom}$).

Note that in this entire discussion, we are using $e^{G}, e^A, e^B, e^C$ which have units of length, and $e^\phi$ has no units.  Hence, one can see that $Q$ must have units of length, and so $q_{\rm geom}$ has units of $\frac{1}{L^4}$.  This is indeed the proper normalization for a charge density if the potential has units of length, $E\sim q V_2 \mu_{\rm geom}$ is an energy, and so $q\sim\frac{1}{L^4}$.  However, when we go to holographic variables, we will want $\mu$ to have units of energy, and this is given by $\frac{\mu_{\rm geom}}{L^2}$ (this is how we map length units in AdS to energy units in the field theory).  The conjugate variable must also map, so we find
\be
q_{\rm geom}L^2= n
\ee
where $n$ is now the field theory number density.  As expected this has units of $L^{-2}$, a unitless number divided by two-volume.

\subsection{Perturbation theory at the horizon}

Near the horizon, we expect a linear zero in $e^{2A}$, a simple pole in $e^{2C}$, and that $\mf$ and the dilaton $e^{\phi}$ go to constants as well.  In what follows, we will fix coordinate transformations by changing to ``entropy gauge'' $e^{2B}=L^2 r^2$, such that the horizon radius $r_h$ is a direct measure of the entropy density.  Therefore, we expand as follows
\bea
A&=&\ln\left(Lr \left(a_0(r-r_h)^{\frac12}+a_0a_1(r-r_h)^{\frac32}+\cdots \right)\right) \\
B&=&\ln\left(Lr\right) \\
C&=&\ln\left(\frac{L}{r}\left(c_0(r-r_h)^{-\frac12}+c_1(r-r_h)^{\frac12}+\cdots\right)\right) \\
G&=&\ln\left(a_0\left(g_h+g_0(r-r_h)+\cdots\right)\right)\\
\phi&=&\ln\left(\Phi_h+p_1(r-r_h)+\cdots\right).
\eea
As expected, we find a constraint on the initial conditions
\be
c_0=\frac{r_h^{\frac12}}{\sqrt{3-\frac{Q^2\Phi_h^{-2\alpha}}{L^2r_h^4}}}.
\ee
Other than this, the equations of motion simply furnish the conserved quantities
\bea
D_0&=&\frac{L^2r_h^4\alpha a_0}{c_0} \\
P_0&=&\alpha Q a_0 g_h=\frac{g_0g_h r_h^2 a_0 \Phi_h^{2\alpha}}{c_0}\\
Q&=&\frac{g_0 r_h^2 \Phi_h^{2\alpha}}{c_0}.
\eea
As promised, only $P_0$ is gauge dependent, depending explicitly on $g_h$, the constant mode in $e^G$.

From the expansion, we can also read the temperature
\be
T=\frac{r_h^2 a_0}{4\pi c_0 L}.
\ee
where we have restored units with $L$ (we will sometimes refer to a unitless temperature $\hat{T}=TL$).  Recall that in the above we have assumed that $a_0$ is picked so that $e^{A}$ asymptotes to $Lr$ with coefficient $1$.  Another way of packaging the same material is to say that $a_0$ is chosen so that $e^{A}$ asymptotes to $a_{\infty}Lr$.  One would then read the temperature as
\be
T=\frac{r_h^2 a_0}{4\pi c_0 L a_{\infty}}.
\ee
This expression is time rescaling invariant, and so can be used in any (time rescaled) gauge.

Finally, we note that the $P_0=0$ gauge has some physical interpretation at this point.  When we consider a Euclideanized time coordinate, the choice $g_h=0$ is necessary to make the one form $e^{G} dt$ well defined on the ``cigar'' geometry.  In some sense this makes $P_0=0$ the preferred gauge.

\subsection{Perturbation theory at $r=\infty$: AdS asymptotics}

Here we expand about $r=\infty$, and require that the solution asymptotes to AdS.  Our perturbative parameter will always be powers of $1/r$.  However, note that the Hamiltonian with non-vanishing charge $Q$ plugged in will be ``zero'' only in the large $r$ limit.  Because of this, we find the set of equations
\bea
&&2e^{A+2B-C}\pa A \pa B+e^{A+2B-C}(\pa B)^2+e^{-A+2B-C+2G+2\alpha \phi}(\pa G)^2 \nn\\
&&\qquad \qquad \qquad \qquad \qquad \qquad-e^{A+2B-C}(\pa \phi)^2+e^{A+2B+C}\Lambda=0 \\
&&2e^{A+2B-C}\left(2\pa \phi+\alpha(\pa A-\pa B)\right)= D_0 \\
&&e^{A+2B-C}\pa \phi+\alpha Q e^{G}= P_0\\
&&e^{-A+2B-C+G+2\alpha\phi}\pa G= Q
\eea
easier to expand.  We parameterize our expansion as
\bea
&& A(r)=\ln(Lr)+A_1(r), \qquad B(r)=\ln(Lr) \\
&& C(r)=\ln(L/r)+C_1(r), \qquad G(r)= \ln(g_b)+G_1(r) \\
&& \phi(r)=\ln(\Phi_b) + \phi_1(r)
\eea
where we have taken that functions with a subscript are perturbative; they fall off at large $r$.  Further, we note that in the above we have already restricted to the ``entropy gauge'' where $r$ directly measures the horizon area.

It is a straightforward matter to plug in the above functions and simply integrate the equations to find a solution.  While doing so, integration constants are introduced that shift the boundary value of $g_b$ and $\Phi_b$, and so we absorb these integration constants into the definition of $g_b$ and $\Phi_b$.  We find
\bea
A_1(r) &=& A_b-\frac{D_0+4\alpha Q g_b -4P_0}{6 \alpha L^2 r^3}+\frac{Q^2}{2 \Phi_b^{2\alpha} L^2 r^4} \\
C_1(r) &=& \frac{D_0+4\alpha Q g_b -4P_0}{6 \alpha L^2 r^3}-\frac{2 Q^2}{3 \Phi_b^{2\alpha} L^2 r^4} \\
G_1(r) &=& -\frac{Q}{g_b \Phi_b^{2\alpha}r} \\
\phi_1(r) &=& \frac{4\alpha Q g_b -4P_0}{12 L^2 r^3}-\frac{Q^2\alpha}{4 \Phi_b^{2\alpha} L^2 r^4}
\eea
We note that in the above, we may remove $A_b$ at the cost of changing the definition of $g_b$ by a rescaling of the time coordinate.  We do so, and so we effectively set $A_b=0$.  At the horizon, this is interpreted as taking a value of $a_0$ such that the function $e^{A}$ asymptotes to $Lr$ with coefficient $1$ (rather than any other constant).  We will see later that this corresponds to picking an appropriate value of $D_0$.

In fact one can read one interesting result directly from the above.  We expect the ``energy mode'' to be determined by the $\frac{1}{r^3}$ mode in $A_1$ or $C_1$.  Therefore, the above gives us a very important piece of information.  We first set $P_0=0$ so that the horizon value of $e^{G}$ is 0 (and so defines a good one-form).  Comparing to the horizon data, we see that $D_0 \propto Ts$ and $Q g_b\propto n \mu$, where $n$ is the number density and $\mu$ is the chemical potential in the field theory.  This gives us that the energy density $E=C_1 Ts + C_2 \mu n$ with $C_1$ and $C_2$ being constants. We will have more to say about this in section 3.

There is of course the black brane solution in pure AdS space given by
\bea
&& A(r)=\ln\left(Lr \sqrt{1-\left(\frac{r_h}{r}\right)^3}\right), \qquad B(r)=\ln(Lr) \\
&& C(r)=\ln\left(\frac{L}{r\sqrt{1-\left(\frac{r_h}{r}\right)^3}}\right) \qquad G(r)=\ln(g_b). \\
&& \phi(r)=\ln({\Phi_b}).
\eea
This solution has $Q=0$, $P_0=0$ and $D_0=3L^2\alpha r_h^3$.  Note that in the euclidean version of the above background that $e^{G(r)}=0$ is the only constant value allowed so that no singularity occurs at the tip of the ``cigar'' geometry.

\subsection{Perturbation theory about Lifshitz asymptotics}

We begin by writing down the solution for the Lifshitz type solution.  The equations read
\bea
&& A(r)=\ln(a_L L_L r^z), \qquad B(r)=\ln(L r), \qquad G(r)=\ln(a_L g_L r^{z+2}+\hat{g}_L) \nn \\
&& 2\alpha \phi(r)= \ln(r^{-4} \Phi_L), \qquad C(r)=\ln\left(\frac{L_L}{r}\right).
\eea
Above we have used the two length scales $L$ (appearing in $B(r)$) and $L_L$ (appearing in $C(r)$).  The scale $L$ is given in terms of the cosmological constant $\Lambda$ in the same way as other sections, so that the coordinate gauge choice for $B(r)$ is the same.  All of the constants in the above solution are given by
\bea
&& L_L^2=\frac{(z+2)(z+1)}{6}L^2 \\
&& \Lambda=-\frac{3}{L^2} \\
&& \alpha=\frac{2}{\sqrt{z-1}} \leftrightarrow z=\frac{\alpha^2+4}{\alpha^2}\\
&& \Phi_L=\frac{Q^2(z+1)}{3L^2(z-1)} \\
&& g_L=\frac{L^2(z-1)}{2 Q} \\
&& P_0=2\frac{Q \hat{g}_L}{\sqrt{z-1}}
\eea
and in the above background, the conserved quantity $D_0=0$.  One may read the above in the same way as we read the black horizon functions: the quantity $L_L$ is not an independent parameter, but depends on other constants.  However, this time it does not depend on other ``horizon'' information, only on information from the Lagrangian $\alpha, L$.  Note also that we could choose a $U(1)$ gauge where the constant part of ${\mathcal A}_t$ is zero,
\be
\hat{g}_L=0,
\ee
at the ``horizon'' and we do this henceforth, i.e. the background we are dealing with has $\hat{g}_L=0$ and so $P_0=0$ to leading order.

We write also the conserved quantity $Q$ in terms of the ``horizon'' values of the fields as we did before
\be
Q=\frac{6 g_L \Phi_L}{z+1}.
\ee

We now expand as before, taking
\bea
&& A(r)=\ln(a_L L_L r^z)+A_1(r), \qquad B(r)=\ln(L r), \qquad G(r)=\ln(a_L g_L r^{z+2})+G_1(r) \nn \\
&& 2\alpha \phi(r)= \ln(r^{-4} \Phi_L)+2\alpha \phi_1(r), \qquad C(r)=\ln\left(\frac{L_L}{r}\right)+C_1(r)
\eea
assuming that the perturbing functions with $_1$ subscripts are small. One may solve all of the equations via a single power law $r^n$ Ansatz because all terms come out homogeneous in powers of $r$.  Doing so, we find solutions for the perturbed equations to be
\bea
&& A_1(r)= \mc C_1 r^{-\frac{z}{2}-1+\frac{\gamma}{2}}+\mc C_2 r^{-\frac{z}{2}-1-\frac{\gamma}{2}} -\frac{D_0 \sqrt{z-1}}{4(z+1)L^2 a_L}r^{-z-2} +\hat{A}_1 \\
&& C_1(r)=\mc C_1 r^{-\frac{z}{2}-1+\frac{\gamma}{2}}+\mc C_2 r^{-\frac{z}{2}-1-\frac{\gamma}{2}} +\frac{D_0 \sqrt{z-1}}{4(z+1)L^2 a_L}r^{-z-2} \\
&&G_1= \mc C_1\frac{\left(-\frac{z}{2}-1+\frac{\gamma}{2}\right)}{z-1} r^{-\frac{z}{2}-1+\frac{\gamma}{2}}+\mc C_2\frac{\left(-\frac{z}{2}-1-\frac{\gamma}{2}\right)}{z-1} r^{-\frac{z}{2}-1-\frac{\gamma}{2}} \nn \\
&&\qquad \qquad \qquad \qquad \qquad+\frac{D_0(z-1)+2P_0(-z-2)}{2\sqrt{z-1}a_L (z+2) L^2}r^{-z-2}+\hat{A}_1 \nn \\
&&\phi_1(r)=-\mc C_1 \frac{1}{\sqrt{z-1}} r^{-\frac{z}{2}-1+\frac{\gamma}{2}}-\mc C_2 \frac{1}{\sqrt{z-1}}r^{-\frac{z}{2}-1-\frac{\gamma}{2}}
\eea
where we have defined the useful constant
\be
\gamma=\sqrt{(z+2)(9z+10)}.
\ee
The modes above were already found in \cite{Goldstein:2009cv}.

A few comments are in order.  First, the constant $\hat{A}_1$ appearing above may be removed by time rescaling, and amounts to a shift in the definition of $a_0$.  Henceforth we take $\hat{A}_1=0$.  Further, the terms proportional to $D_0$ are related to the mass parameter given in \cite{Taylor:2008tg,Gubser:2009qt,Goldstein:2009cv}.  Further, the last two terms in $G_1$ may be set to zero via an appropriate gauge choice, as the term $r^{-2-z}$ in $G_1$ is a pure $U(1)$ gauge transformation.

Note that if we want to add an IR ($r\rightarrow 0$) irrelevant perturbation, the only mode that available is $r^{-\frac{z}{2}-1+\frac{\gamma}{2}}$ \cite{Goldstein:2009cv}.  Further, it should be noted that this term does not affect the values of $D_0$ or $P_0$.  Therefore, one can imagine flowing from the $P_0=0,D_0=0$ Lifshitz fixed point (with $\hat{g}_L=0$) to the $D_0=0, P_0=0$ AdS fixed point.

Finally, we write down the exact black brane solution in \cite{Taylor:2008tg,Gubser:2009qt,Goldstein:2009cv}
\bea
&& A(r)=\ln\left(a_L L_L r^z \sqrt{1-\left(\frac{r_h}{r}\right)^{2+z}}\right) \nn \\
&& B(r)=\ln(Lr) \nn \\
&& C(r)=\ln\left(\frac{L_L}{r \sqrt{1-\left(\frac{r_h}{r}\right)^{2+z}}}\right) \nn \\
&& G(r)= \ln\left(a_L g_L (r^{2+z}-r_h^{2+z})\right) \nn \\
&& 2\alpha \phi(r)= \ln\left(r^{-4}\Phi_L\right).
\eea
All constants are as before, and $r_h$ is the location of the horizon.  This background has
\be
P_0=0, \qquad
D_0=\frac{2a_L L^2 r_h^{2+z}(2+z)}{\sqrt{z-1}}
\ee
($P_0$ was set to 0 by construction, see the section about the perturbation theory about the horizon).  This allows that all of the asymptotic functions found above may be removed by redefinition of $r_h$, or $U(1)$ gauge, or time rescaling except for those multiplied by $\mc C_1$ and $\mc C_2$.

\subsection{Setup for numeric integration}

From the earlier discussions, we have found that $P_0=0$ is a convenient $U(1)$ gauge choice.  We will want to take a particular coordinate gauge condition as well, and above we have always taken
\be
B=\ln\left(Lr\right).
\ee
This completely fixes the coordinate gauge.  Further, we will find it useful to reexpress the differential equations in terms of ``correction'' functions as
\bea
A&=&\ln\left(Lr\right)+A_1(r) \\
B&=&\ln\left(Lr\right) \\
C&=&\ln\left(\frac{L}{r}\right)+C_1(r) \\
G&=&\ln\left(L\right)+G_1(r)\\
\phi&=&\phi.
\eea
What we have essentially done here is to pull off all of the units from the functions, and so the resulting differential equations will be unitless if we identify the correct units for $P_0, D_0$ and $Q$ and use $L$ to construct unitless quantities.  The correct identification is
\bea
&& D_0=\hat{D}_0 L^2 \\
&& P_0=\hat{P}_0 L^2=0 \\
&& Q=\hat{Q} L \\
\eea
where the hatted quantities are unitless.  Inserting these into the above differential equations, we see that one has $4$ first order differential equations for $3$ functions.  Therefore, one combination must be algebraic.  We find this combination and solve for $\phi$ to find
\bea
&& e^{2\alpha \phi}=\label{phialg} \\
&& \frac{\alpha \Qh^2 r^2 e^{A_1+C_1}}{3\alpha r^6 \left(e^{A_1+C_1}-e^{A_1-C_1}\right)+\alpha e^{-A_1 + C_1}\left(\alpha \Qh e^{G_1}\right)^2-4r^3\left(\alpha \Qh e^{G_1}\right)-r^3 \Dh_0}\nn
\eea
The remaining differential equations become
\bea
&& \kern -2 em \pa e^{A_1} - \frac12 \frac{e^{C_1}\left(4\alpha \Qh e^{G_1}+\Dh_0\right)}{r^4 \alpha}=0 \\
&& \kern -2 em \pa e^{C_1} - \frac12 \frac{\left(e^{C_1}\right)^2\left(2\alpha^3 e^{C_1}\Qh^2\left(e^{G_1}\right)^2-4\alpha \Qh e^{G_1} e^{A_1}r^3-\Dh_0 e^{A_1}r^3\right)}{\alpha \left(e^{A_1}\right)^2 r^7}=0 \\
&& \kern -2 em\pa e^{G_1} \nn \\
&& \kern -2 em-\frac{3\alpha r^6 \left(e^{A_1}\right)^2\left(\left(e^{C_1}\right)^2-1\right)
+\alpha^3 \Qh^2 \left(e^{C_1}e^{G_1}\right)^2
-r^3e^{C_1} e^{A_1}\left(4\alpha \Qh e^{G_1}+\Dh_0\right)}{\Qh \alpha r^4 e^{C_1} e^{A_1}} \nn \\
&& \kern 25em =0
\eea

We now remark on several features of the differential equations.  First off, there is the global symmetry already mentioned $(A,B,C,\phi,G)\rightarrow (A,B,C,\phi+\delta_2,G-\alpha\delta_2)$.  This symmetry has the effect of rescaling the charge $Q$ in the differential equations by $Q\rightarrow e^{\delta_2 \alpha}Q$, correctly reproducing the transformation property in equation (\ref{phialg}).  Hence, using this global symmetry, one may in fact set $\Qh$ to be any value one wishes, knowing that all solutions are related to this one by the global symmetry \footnote{Similar considerations were noted in \cite{Goldstein:2009cv} for extremal solutions.  However, here we have 2 quantities affected by the symmetry of AdS rescaling the coordinates $(t,x_i,r)\rightarrow (\lambda t,\lambda x_i,\lambda^{-1} r)$, $Q$ and $D_0$ (and $P_0$, but this is pure gauge).  Therefore, rather than having all solutions related to one solution, as the authors of \cite{Goldstein:2009cv} had for the extremal solutions, we expect to have a one parameter family of solutions that are distinct.  This is essentially getting at the fact that the thermodynamics should be sensitive only to $\frac{\mu}{T}$, and this is our one parameter family.}.  One can go further with this line of reasoning.  Note that if we take a time rescaling, this has the effect of taking $(A,B,C,\phi,G)\rightarrow (A+\delta,B,C,\phi,G+\delta)$.  While this is {\underline{not}} a global symmetry of the reduced action, it simply rescales the action, and so maps solutions to solutions (this is an example of a Lie point symmetry).  Further, one can see that only $D_0$ and $P_0$ rescale under this transformation.  Therefore, as long as $D_0$ is not zero, we may scale this to any quantity we wish.  {\it Therefore, to efficiently parameterize the horizon data that we start with, we take the inputs $\hat{D}_0=3 \alpha r_h^3$ (which, given $r_h$ is always possible to set via time rescaling), and $\hat{Q}=1$ (which is always possible using the global symmetry).}  We note, however, that this will lead to fields not being normalized at infinity.  Then, the rescalings used to normalize them are written easily in terms of the asymptotic values of the fields which is the output of the numeric integration.  This then allows us to conclude what the ``correct'' value of $\hat{Q}$ and $\hat{D}_0$ should have been.

Using $\hat{D}_0=3 \alpha r_h^3$ and $\hat{Q}=1$ as inputs, we numerically integrate given initial data $c_0$ and $r_h$.  Conveniently, $c_0$ only has a window in which it is defined (given the black brane in AdS and the black brane in Lifshitz solutions)
\be
\sqrt{\frac{r_h}{3}}\leq c_0 \leq \frac{\sqrt{\alpha^2+2}}{\alpha}\sqrt{\frac{r_h}{3}}.
\ee
Physically, when $c_0$ approaches $\sqrt{\frac{r_h}{3}}$, we get the usual black brane in AdS.  This should be a zero charge, zero chemical potential solution.  As $c_0$ increases to $\frac{\sqrt{\alpha^2+2}}{\alpha}\sqrt{\frac{r_h}{3}}$, it approaches a Lifshitz asymptotic solution, and such a solution has a divergent potential\footnote{One can see this as the combined function $e^{\alpha \phi} e^{G}$ diverging in the Lifshitz solutions.  This combination is a global symmetry invariant and is saying that $\mu$, as we have defined it, is becoming large.} and finite temperature, and so should be considered a $\frac{\mu}{T}\rightarrow \infty$ solution.  We always include an AdS completion, and fix the value of $\mu$, and therefore the above limit should be an extremal limit.  Such a solution would simply be the spacetime with the $r\rightarrow 0$ limit being the Lifshitz solution, and the $r\rightarrow \infty$ limit being AdS numerically studied in \cite{Goldstein:2009cv}.  These expectations will bear out in our numeric analysis.

For completeness, we include the seed functions used to supply the initial conditions at (actually near) $r=r_h$,
\bea
&& e^{A_1}= a_0 \left( \sqrt{r-r_h}+ a_1 \left(r-r_h\right)^{\frac{3}{2}}+ \cdots \right) \\
&& e^{C_1}= c_0 \frac{1}{\sqrt{r-r_h}} + c_1 \sqrt{r-r_h} + \cdots\\
&& e^{G_1}= a_0 \left( g_0(r-r_h) + g_1 (r-r_h)^2+ \cdots \right)
\eea
with
\bea
&& a_0= \frac{c_0 \hat{D}_0}{\alpha r_h^4},\quad  a_1=\frac{9\alpha^2 c_0^4+6c_0^2 r_h(2-\alpha^2)+(\alpha^2-8)r_h^2}{4 r_h^3}\\
&& c_0= c_0, \quad c_1=\frac{c_0\left(27 \alpha^2 c_0^4 - 6 c_0^2 r_h(2+3\alpha^2)+r_h^2(8+3\alpha^2)\right)}{4 r_h^3}\\
&& g_0= \frac{(3c_0^2-r_h)r_h^2}{c_0 \hat{Q}}, \quad g_1=\frac{ r_h^3(2+\alpha^2)-3c_0^2r_h^2(2+\alpha^2)-9c_0^4r_h\alpha^2+27 c_0^6\alpha^2}{2c_0 r_h \hat{Q}}
\eea

As mentioned above, after performing the numeric integration we see that $e^{A_1}$ does not asymptote to $1$.  Therefore, to get a physically meaningful result, we have to do a time rescaling.  This essentially rescales the value of $\hat{D_0}$ to be the correct value to give the correct asymptotic.  This is the first piece of output: we get the physical value of $D_0$ that should be associated with the solution, and this is used to determine part of the energy density.

However, now we run into a small puzzle.  The physical value of $e^{G_1}$ is not strictly determined.  We may use the global symmetry $(A,B,C,\phi,G)\rightarrow (A,B,C,\phi+\delta_2,G-\alpha\delta_2)$ to rescale this value to any value we wish.  Because this is a global symmetry that does not involve the metric, the stress energy tensor (and therefore the geometry), \underline{is not} determined by this number.  In fact, only global symmetry invariants can determine anything in the geometry (this was also noticed in \cite{Goldstein:2009cv} for extremal solutions).  We wish to consider $\mu$ as a scale in the theory, i.e. it is the scale at which new particles can be added/excited, and so we expect this to correspond to some scale in AdS, and so must be a global symmetry invariant.

One way to fix this ambiguity is to note that the gauge coupling at infinity is the asymptotic value of $e^{2 \alpha \phi}$ (which also transforms under the global symmetry).  Hence, we would like the gauge kinetic term to go to a canonical value.  This can be arranged by using the global symmetry to scale $e^{2 \alpha \phi}|_{r=\infty}=1$.  From the beginning we should have expected such a statement: only fixing both the asymptotic value of $e^{\alpha \phi}$ {\it and} the asymptotic value of $e^{G_1}$ will determine the geometry.  We had traded this for $e^{G_1}$ and $\hat{Q}$ in the above discussions.  Therefore, one can read the output value of $\mu$ from the numeric integration as
\be
\hat{\mu}= \frac{e^{G_1} e^{\alpha \phi}}{e^{A_1}}\Bigg|_{r=\infty}.
\ee
The factor of $e^{A_1}$ in the denominator accounts for the fact that $e^{G_1}$ transforms under the time rescaling used to set the asymptotic value of $e^{A_1}$ to be $1$, and the factor of $e^{\alpha \phi}$ in the numerator is there because of using the global symmetry to set $e^{\alpha \phi}=1$.  Let us assume that $e^{A_1}$ goes to 1 at infinity.  Then the above simply states that what we have done is take some bare quantities $q_{\rm geom}$ and $\mu_{\rm geom}$ and combined them into the global symmetry invariant $q_{\rm geom}\mu_{\rm geom}= \left(e^{\alpha \phi}|_{r=\infty}q_{\rm geom}\right)\left(e^{-\alpha \phi}|_{r=\infty}\mu_{\rm geom}\right)$.  The second expression is made out of global symmetry invariants, and so can define scales in the geometry.  These are what we use to define the physical chemical potential $\mu_{\rm geom}$, and why we will find $e^{\alpha \phi}|_{r=\infty}=1$ a convenient representative from all solutions related by the global symmetry.

As with all other formulae, the above has units restored using only $L$ via $\mu_{\rm geom}=L \hat{\mu}$, and this defines the length scale in the geometry, and so $\mu=\frac{\hat{\mu}}{L}$ defines an energy scale in the field theory.  One should further note that $\hat{Q}$ has changed under this global symmetry rescaling and so becomes an output of the numeric analysis as $\hat{Q}=\frac{\hat{Q}_{\rm input}}{e^{\alpha \phi}}\big|_{r=\infty}=\frac{1}{e^{\alpha \phi}}\big|_{r=\infty}$.

Schematically, what we have then is that $(c_0, r_h, \alpha; \hat{Q}_{\rm input}=1, \hat{D}_{0, {\rm input}}(r_h,\alpha) )$ is input, and this gets mapped to $(T, \mu, \hat{D}_0, \hat{Q})$, which are determined by the asymptotic values of the fields.

We expect all relevant physical information is contained in $\frac{\mu}{T}$, given the scaling invariance of AdS (see the next section).  One way to scan through such values is to start with an $r_h$ and scan through values of $c_0$ in the window of allowed values to find a fixed value of $\mu$.  Then we may scan through $r_h$ for different values.  This process is fixing $\mu$ and scanning through $r_h$ to see what the output values of $T$ are.  If $T$ becomes multiple valued, then we expect a phase transition, and if it is monotonic, there is no phase transition.  Further, we expect that as $r_h \ll \hat{\mu}$ we will get Lifshitz like behavior, and for $r_h \gg \hat{\mu}$ we will get AdS like behavior.  This is essentially the limits $T\ll \mu$ and $T \gg \mu$ in the field theory.

We now turn to the results of the numeric analysis we have described above, and our result relating ${\mc{E}}$ to $Ts$ and $\mu n$.

\section{Discussion and results}

\begin{figure}
\centering
\includegraphics[width=0.4\textwidth]{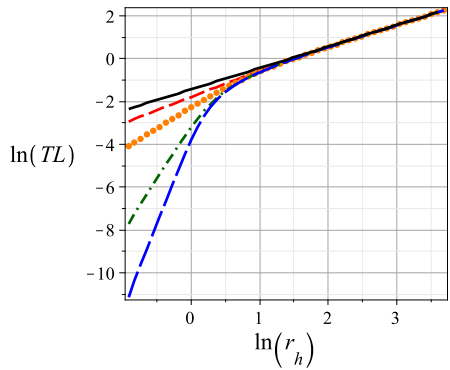}
\caption{Above we have plotted $\ln(TL)$ as a function of $\ln(r_h)$ for fixed $\hat{\mu}=2$.  The different graphs correspond to different values of $z$: $z=1.25 \; (\alpha=4)$ (red dashed curve), $z=2 \;(\alpha=2)$ (orange dotted curve), $z=5 \;(\alpha=1)$ (green dash-dotted curve), $z=\frac{73}{9}= 8.\bar{1} \;(\alpha=0.75)$ (blue long-dashed curve).  The solid black curve is a plot for the pure AdS black brane $\ln(TL)=\ln\left(\frac{3r_h}{4\pi}\right)$, the asymptotic value of all graphs in the $\ln(r_h) \rightarrow \infty$ limit.  The slopes of the graphs approach $z$ as $\ln(r_h)\rightarrow -\infty$.}
\label{Tgraphs}
\end{figure}

The first point we wish to address is whether there is a phase transition.  More concretely, we wish to determine whether there is more than one black brane solution given a value of $\hat{\mu}$ and $\hat{T}$.  We can do this by setting $\hat{\mu}=2$ (the value is arbitrary) and then seeing whether the temperature is a single or multiple valued function of $r_h$.  We show log-log plots\footnote{Maple worksheets available upon request} of the temperature $\hat{T}=TL$ as a function of $r_h$ in figure \ref{Tgraphs}.
We find that we always get monotonic behavior for a wide range of values of $\alpha$.  Hence there is no phase transition associated with going from $T\ll \mu$ to $T \gg \mu$.  Further, we see that we get the correct asymptotic behaviors on both ends of the graphs with $\hat{T}\propto r_h$ for $r_h \gg 2$  and $\hat{T}\propto r_h^z$ for $r_h \ll 2$ where $z=\frac{\alpha^2+4}{\alpha^2}$ as before.  Note that the scale of the changeover occurs approximately at $\ln(r_h)=\ln(2)=\ln(\hat{\mu})$.  This justifies the association of $L\hat{\mu}$ with a scale in AdS.

\begin{figure}[ht!]
\centering
\subfloat[$r_h=0.4, \; \hat{\mu}=2$]{\label{metsmall}
    \includegraphics[width=0.3\textwidth,angle=0]{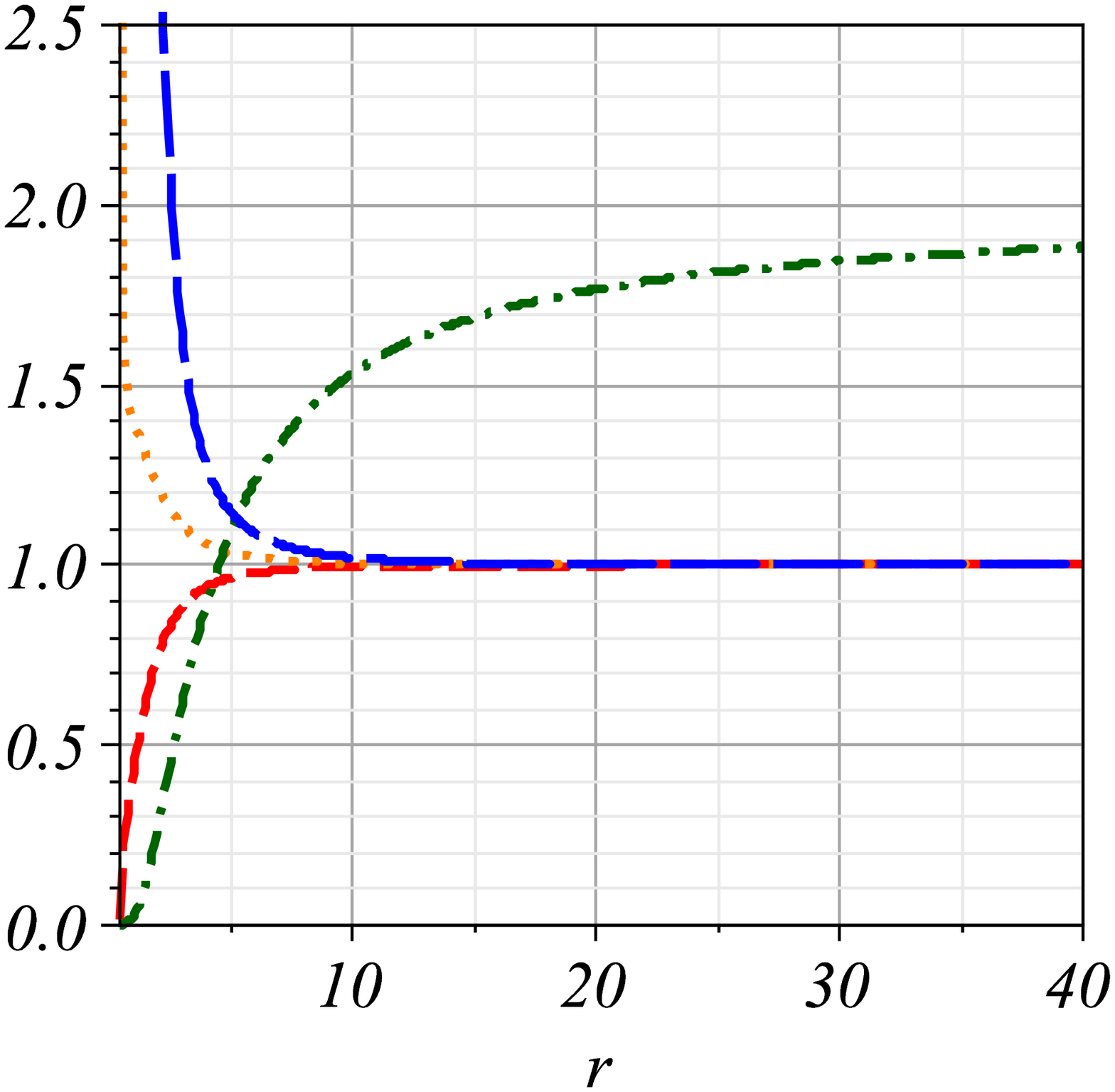}}
\subfloat[$r_h=0.4, \; \hat{\mu}=2$ asym.]{\label{metsmallasym}
     \includegraphics[width=0.3\textwidth,angle=0]{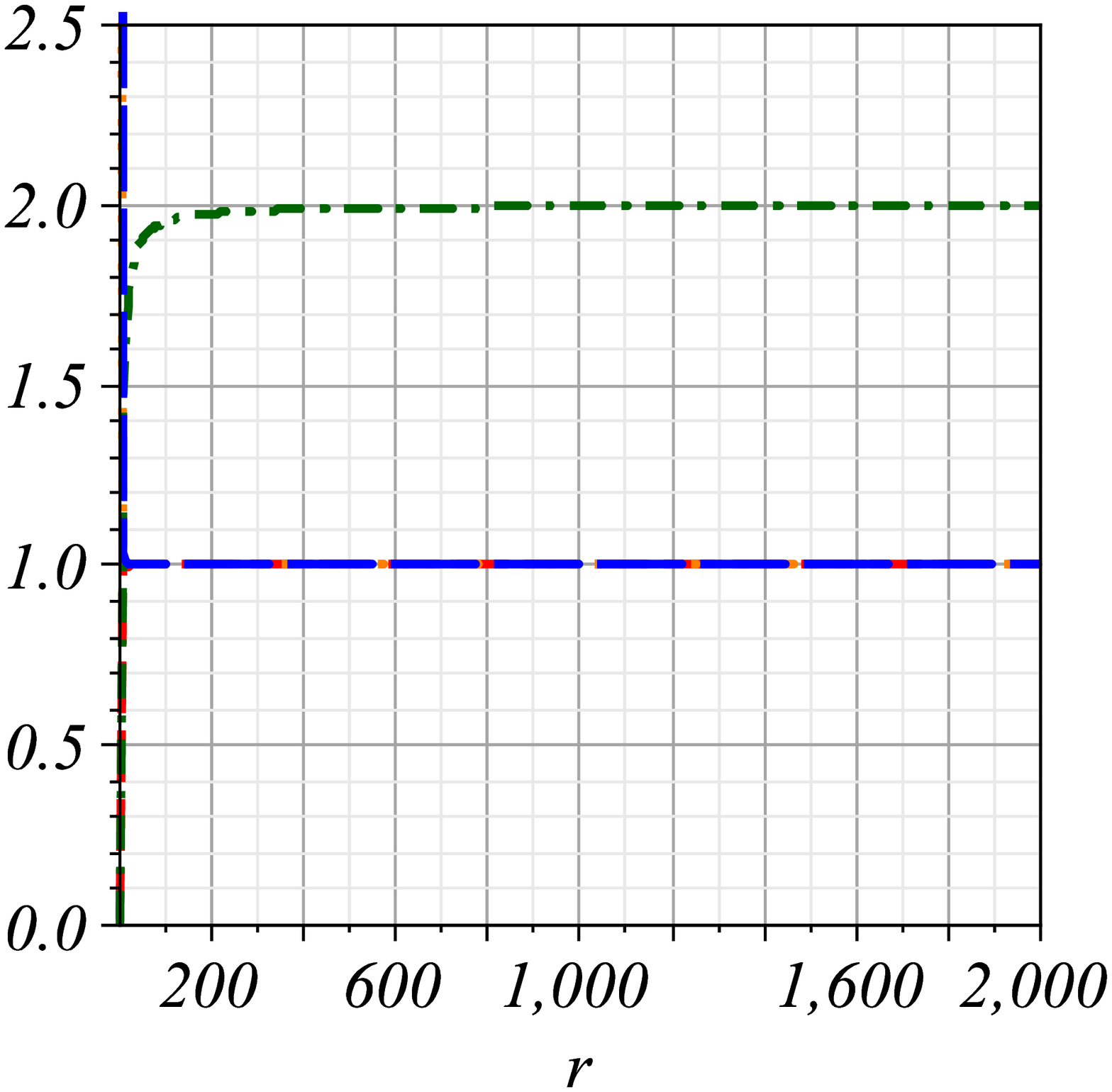}}\\
\subfloat[$r_h=40, \;\hat{\mu}=2$]{\label{metbig}
    \includegraphics[width=0.3\textwidth,angle=0]{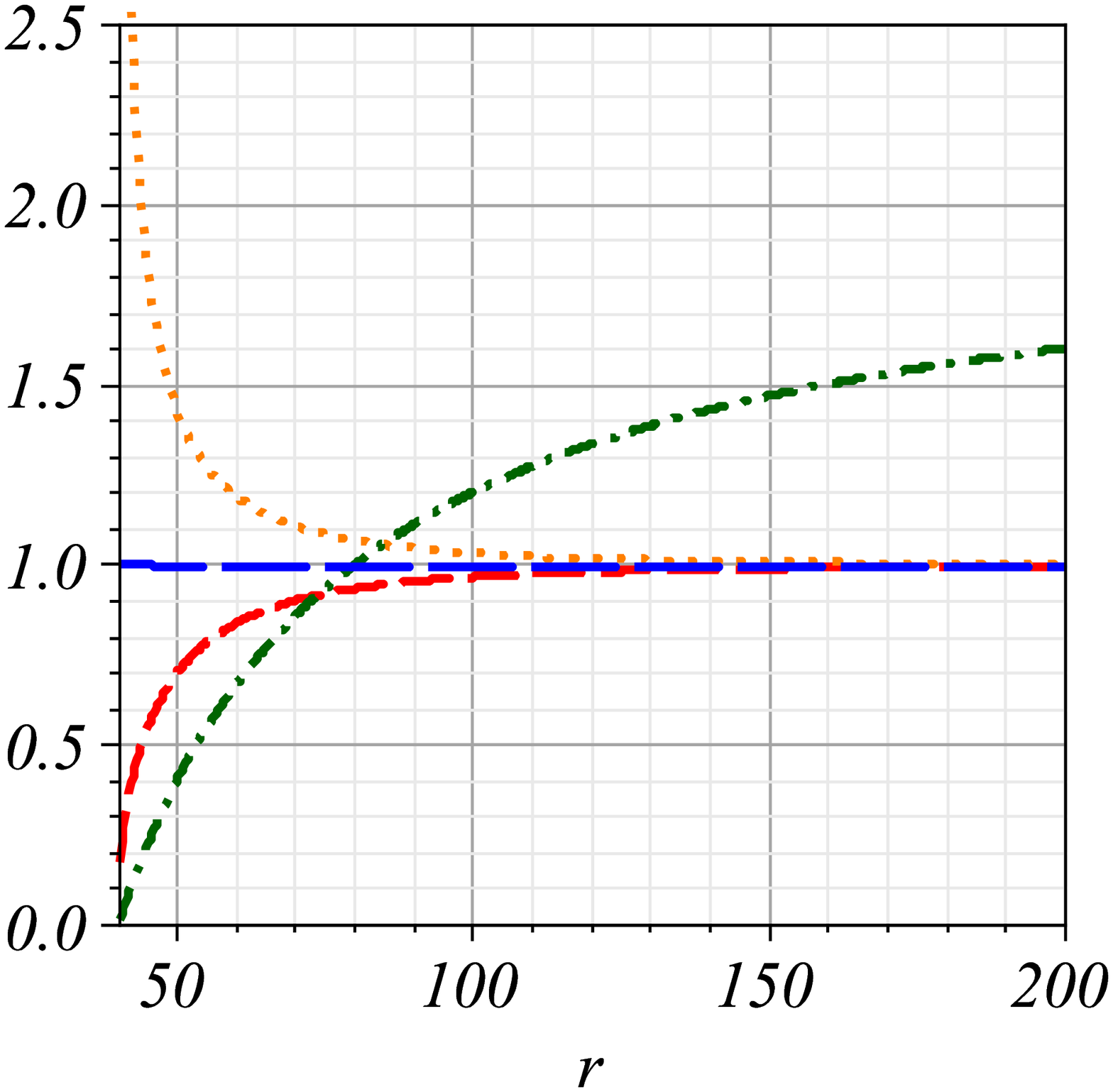}}
\subfloat[$r_h=40,\; \hat{\mu}=2$ asym.]{\label{metbigasym}
     \includegraphics[width=0.3\textwidth,angle=0]{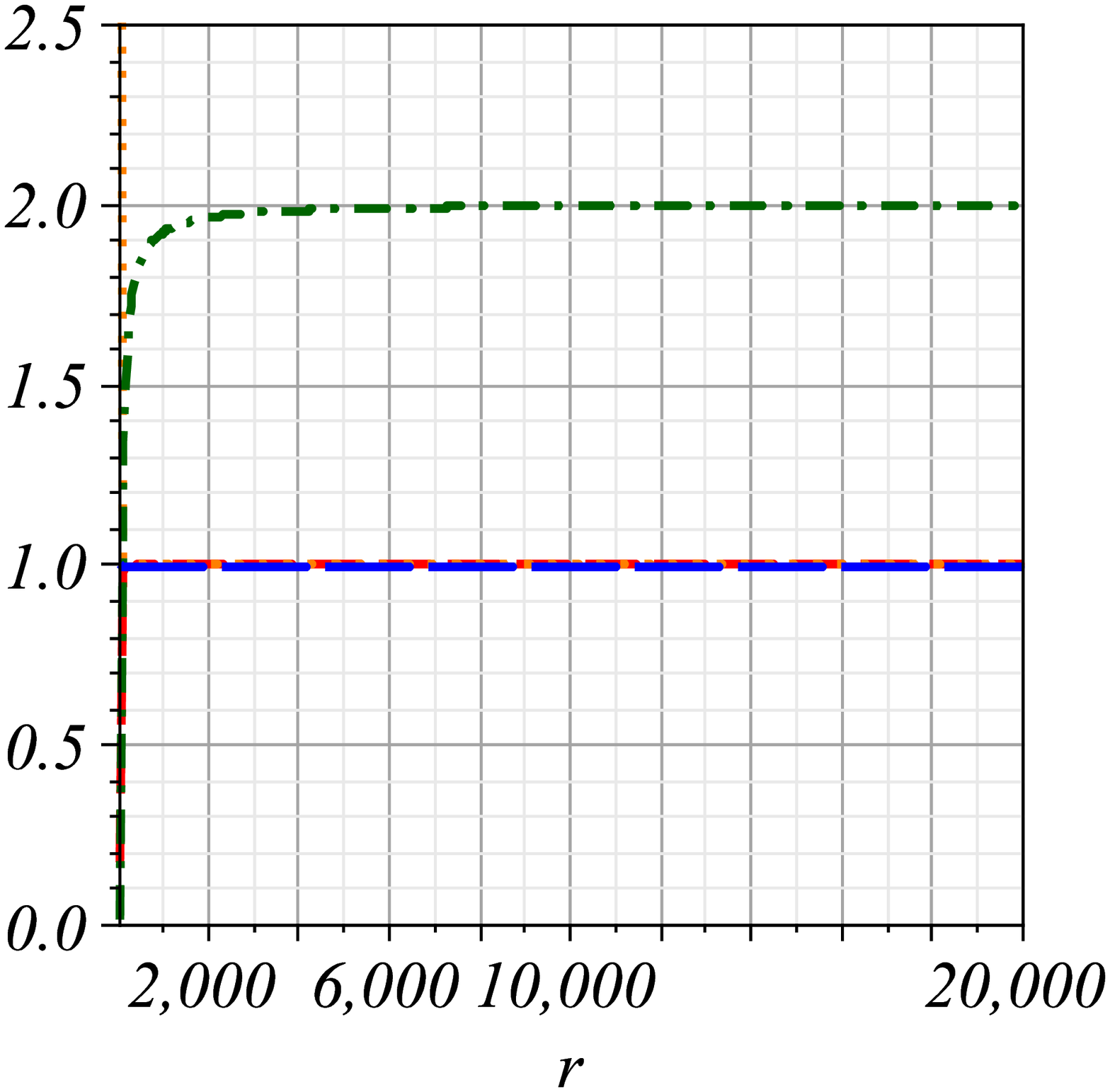}}
\caption{ The above plots depict the metric functions and fields as a function of $r$, the top plots for a small value of  $r_h=0.4$, and the bottom plots for a large value of $r_h=40$.  The plots show $e^{A_1}$ (red dashed curve), $e^{C_1}$ (orange dotted curve), $e^{G_1}$ (green dash-dotted curve) and $e^{2\alpha \phi}$ (blue long-dashed curve).  Plots (b) and (d) simply show that the asymptotics are correct with the convention that $e^{2\alpha \phi}$ is set to $1$ at the boundary.}
\label{metsmallall}
\end{figure}

Next, we graph the functions $e^{A_1}, e^{G_1}, e^{C_1}, e^{2\alpha \phi}$ for $\alpha=2$ after adjusting $\hat{D}_0$ and $\hat{Q}$ to give the correct asymptotic values of $e^{A_1}, e^{G_1}, e^{C_1}, e^{2\alpha \phi}$.  We plot solutions for a small value of $r_h$  and a large value of $r_h$ in figure \ref{metsmallall}.
We see no qualitative difference in behavior in that $e^{A_1}, e^{G_1}, e^{C_1}, e^{2\alpha \phi}$ appear to be monotonic regardless of whether $r_h \ll \hat{\mu}$ or $r_h \gg \hat{\mu}$ (we omit plots for $r_h\sim \hat{\mu}$, as these look very similar too).

\begin{figure}
\centering
\includegraphics[width=0.4\textwidth]{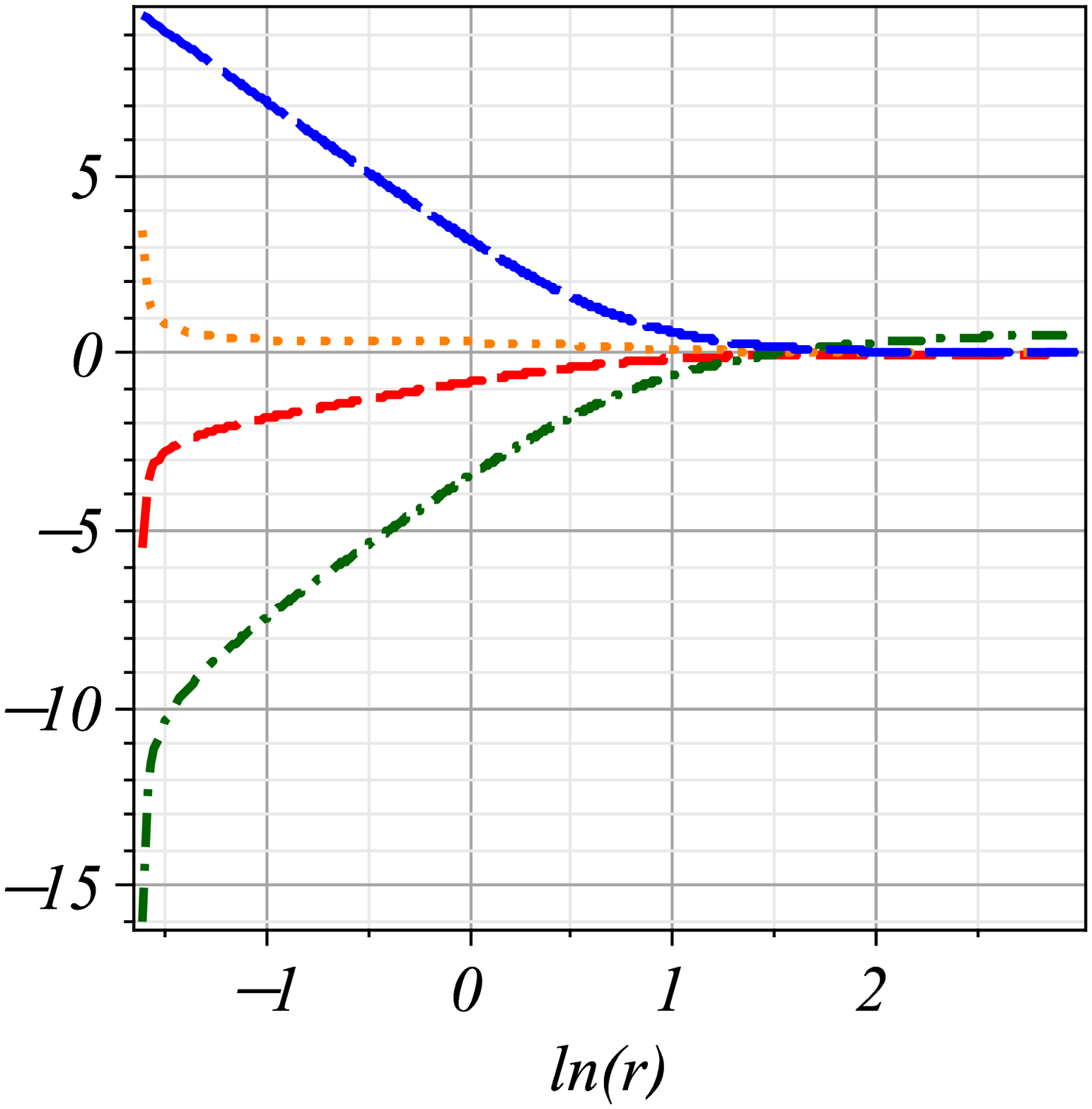}
\caption{Above we have plotted the metric functions $A_1$ (red dashed curve), $C_1$ (orange dotted curve) and the fields $G_1$ (green dash-dotted curve), $2\alpha \phi$ (blue long-dashed curve) as a function of $\ln(r)$ for fixed $\hat{\mu}=2$, and horizon radius $r_h=0.2$.  We see that all the functions have an approximately linear portion in the region $-1<\ln(r)<0$.  The slopes of these linear portions can be shown to approximately reproduce the Lifshitz background with $z=2, (\alpha=2)$.  Further, the asymptotic value of $G_1(r) \rightarrow \ln(2)$ is seen, as opposed to the asymptotic value of all other graphs $\ln(1)=0$.}
\label{supersmall}
\end{figure}

Another way to analyze the solutions is to look at log-log plots of the metric functions and fields for a small value of $r_h$.  For this, we expect to see a section closely approximating the pure Lifshitz spacetime, and indeed this is what we find.  We plot an example of this in figure \ref{supersmall}.

\begin{figure}
\centering
\includegraphics[width=0.4\textwidth]{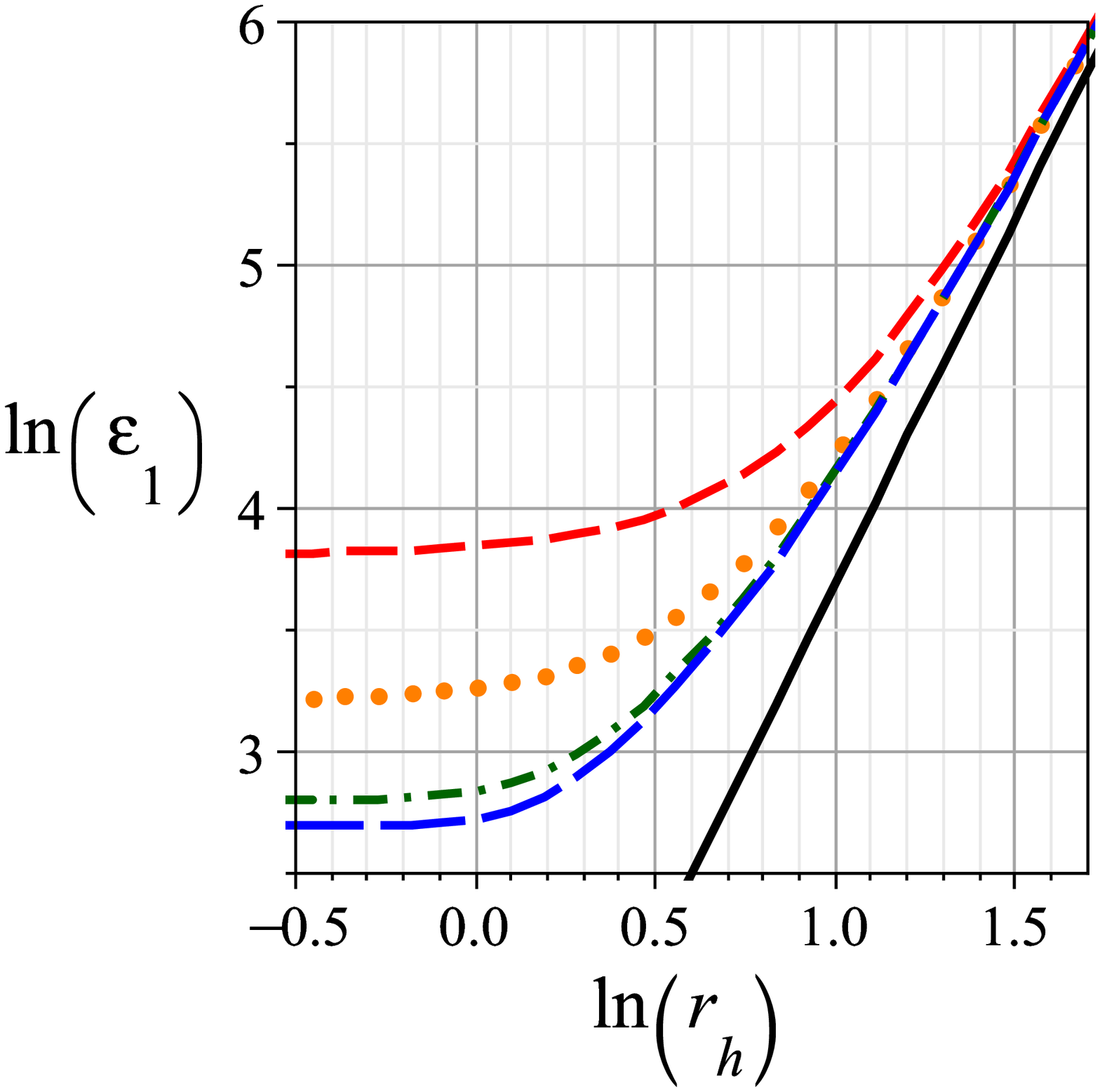}
\caption{Above we have plotted $\ln(\mc E_1)$ as a function of $\ln(r_h)$ for fixed $\hat{\mu}=2$.  The different graphs correspond to different values of $z$: $z=1.25 \; (\alpha=4)$ (red dashed curve), $z=2 \;(\alpha=2)$ (orange dotted curve), $z=5 \;(\alpha=1)$ (green dash-dotted curve), $z=\frac{73}{9}= 8.\bar{1} \;(\alpha=0.75)$ (blue long-dashed curve).  The solid black curve is a plot for the pure AdS black brane $\ln(\mc E_1)=\ln\left(4\frac{3 \alpha r_h^3}{6\alpha}\right)$, the asymptotic value of all graphs in the $\ln(r_h) \rightarrow \infty$ limit.  The slopes of the graphs approach $0$ as $\ln(r_h)\rightarrow -\infty$.  This is merely an indication of a finite energy held in the $U(1)$ gauge field, and may be considered an extremal limit with ``mass $=$ charge'' in the right units.  One can check this by graphing $\frac{Q_{\rm out}}{{\mc E}_1}$ and seeing that it goes to $\frac{6}{16\hat{\mu}}$ (independent of $\alpha$) for $r_h \rightarrow 0$, however we do not display such plots.}
\label{Egraphs}
\end{figure}

Finally, we turn to the matter of the energy density.  From the AdS asymptotics we read that the energy density is
\be
\mc E=\frac{1}{16 \pi G_4 L} 4\frac{D_0+4\alpha Q g_b}{6 \alpha L^2 }
\ee
where we have already set $P_0=0$).  Therefore
\be
\mc E_1=4\frac{\hat{D}_0+4\alpha \hat{Q} g_{1,b}}{6 \alpha }
\ee
is a unitless measure of the energy, and where we have denoted $g_{1,b}=e^{G_1}|_{r=\infty}$ when $e^{2\alpha \phi}|_{r=\infty}=1$.  We plot this in figure \ref{Egraphs}.

Next, we will argue on general grounds that this expression for the energy density should be expected.  First, we recall a few facts,
\bea
&& D_0= L^2 r_h^4 \alpha \frac{a_0}{c_0}, \quad Q=\frac{n 16 \pi G_4 L}{4}\nn \\
&& g_b= L\hat{\mu}=\mu L^2,\quad T=\frac{r_h^2}{4\pi L} \frac{a_0}{c_0}, \quad s= \frac{4\pi r_h^2}{16 \pi G_4}.
\eea
Now, in the energy density we have a term $D_0 \sim r_h^4\sim T s$ and a term $Qg_b\sim \mu n$.  Further, $g_b$ is the value of the potential at the boundary (after we fix $e^{\alpha \phi}=1$ at the boundary), and $g_b Q$ is a global symmetry invariant (under shifting of $G$ and $\phi$).  In deriving the above, we have assumed that $e^{A_1}$ asymptotes to $1$, fixing the time rescaling.  Finally, the above observations allows us to write
\be
\mc E=\frac{2}{3}\left(Ts + \mu n\right).
\ee

In fact, we can derive such a relationship using only conformal symmetry.  We note that $T$ and $\mu$ are energies, $n$ and $s$ are densities, and ${\mc E}$ is an energy density.  Hence, we expect that in $d$ spatial dimensions these quantities should scale as
\be
\left({\mc E},T,\mu,s,n\right)\rightarrow \left(\lambda^{d+1} {\mc E}, \lambda T, \lambda \mu, \lambda^{d}s, \lambda^{d}n\right),
\ee
and so the general functional form must be
\bea
{\mc E}=T^{1+d} f\left(\frac{\mu}{T}\right),\quad s= T^{d} g\left(\frac{\mu}{T}\right), \quad n= T^{d} h\left(\frac{\mu}{T}\right).
\eea
Next, we assume the first law of thermodynamics
\be
d \mc E = T ds+ \mu dn.
\ee
We take the above differentials and equate the $d\mu$ coefficient to find
\be
f'=g'+\frac{\mu}{T}h' \label{dmurel}
\ee
and equate the $dT$ coefficient to find
\be
T^d (1+d) f- T^{1+d}\frac{\mu}{T^2} f'= T^d g d - T^d T \frac{\mu}{T^2}g' +  \mu T^{d-1} h d - \mu T^d \frac{\mu}{T^2} h'.
\ee
Using (\ref{dmurel}) this reads
\be
T^{1+d} f=\frac{d}{d+1}\left(T T^d g+ \mu T^d h\right)
\ee
which is identical to
\be
{\mc E}= \frac{d}{d+1}\left(T s+ \mu n\right).\label{generalErelation}
\ee
Further, the relationship (\ref{dmurel}) becomes
\be
\frac{d}{d+1}\left( g' +\frac{\mu}{T} h' + h\right)=g'+\frac{\mu}{T}h'
\ee
or rearranging a bit
\bea
T^{d-1} g'&=& d T^{d-1} h-T^{d-2}\mu h' \nn \\
\pa_\mu s &=& \pa_T n. \label{intcondnew}
\eea
The generalization of the above argument for arbitrary number of particle species is straightforward, and gives
\be
{\mc E}= \frac{d}{d+1}\left(Ts + \mu_i n_i \right), \qquad \pa_{\mu_i} s= \pa_T n_i + \left(\pa_{\mu_j}n_i-\pa_{\mu_i} n_j\right)\frac{\mu_j}{T}.
\ee
Using this, and the Euler relation $E=TS-PV+\mu_i N_i$ we arrive at the familiar ${\mc E}=P d$: the stress-energy tensor is traceless.

For the case at hand, the argument is reversible, i.e. the scaling relation can be obtained from a relation of the sort
\be
{\mathcal E}=C_1 Ts + C_2 \mu n. \label{nicerelation}
\ee
Again, we assume the first law of thermodynamics
\be
d \mc E = T ds+ \mu dn,
\ee
and write all functions in terms of the thermodynamic variables $T, \mu$ and find
\bea
d \mc E &= &(T \pa_T s + \mu \pa_T n)dT+ (T \pa_\mu s + \mu \pa_\mu n)d\mu \nn \\
 &=& \pa_T\left(C_1 Ts + C_2 \mu n\right)dT+\pa_\mu\left(C_1 Ts + C_2 \mu n\right)d\mu.
\eea
Equating differentials allows us to write two first order differential equations
\bea
T \pa_T s + \mu \pa_T n=\pa_T\left(C_1 Ts + C_2 \mu n\right) \\
T \pa_\mu s + \mu \pa_\mu n=\pa_\mu\left(C_1 Ts + C_2 \mu n\right).
\eea
Taking the integrability condition for these two equations ($\pa_\mu$ acting on the first equation minus $\pa_T$ acting on the second) we find that
\be
\pa_\mu s=\pa_T n, \label{intcond}
\ee
which is the same relation as the one above (\ref{intcondnew}).
This allows us to separate the equations and find
\bea
(1-C_1)T \pa_T s + (1-C_2) \mu \pa_\mu s=C_1 s  \\
(1-C_1)T \pa_T n + (1-C_2) \mu \pa_\mu n=C_2 n.
\eea
The above differential equations have the following solutions
\be
s=T^{\frac{C_1}{1-C_1}}\hat{g}\left(\frac{\mu^{1-C_1}}{T^{1-C_2}}\right), \qquad
n=\mu^{\frac{C_2}{1-C_2}}\hat{h}\left(\frac{T^{1-C_2}}{\mu^{1-C_1}}\right), \label{snslns}
\ee
(these functions are related easily to $g$ and $h$ above) and the integrability condition (\ref{intcond}) relates $\hat h$ to $\hat g$ via
\be
\hat h'\left(\frac{T^{1-C_2}}{\mu^{1-C_1}}\right)=\frac{1-C_1}{1-C_2} \frac{T^{\frac{1}{1-C_1}}}{\mu^{\frac{1}{1-C_2}}}\hat g'\left(\frac{\mu^{1-C_1}}{T^{1-C_2}}\right).
\ee

Next, we would like to know what initial set of data specifies the thermodynamics in the entire $(T,\mu)$ plane.  More specifically, we already know that we have a solution for the $\mu=0$ case (this is the uncharged black brane in AdS case), and would like to know if this, or any additional information, can specify the thermodynamics off of this curve.     To do so, we would like to calculate the characteristic equation for paths in $T,\mu$ space, parametrically given by $T(\sigma), \mu(\sigma)$.  The characteristic curves are those that satisfy
\be
(1-C_1)T \frac{d\mu}{d\sigma}-(1-C_2)\mu \frac{d T}{d\sigma}=0.\label{charequation}
\ee
(we note that the characteristic equations are the same for both of our linear PDEs).  The most general solution to these equations, up to reparameterization of $\sigma$, are
\be
T= T_0 \sigma^{1-C_1}, \qquad \mu = \mu_0 \sigma^{1-C_2}.
\ee
These curves are given by $T^{(1-C_2)}\mu^{-(1-C_1)}={\rm constant}$. These are either hyperbola-like curves that do not intersect $T=0$ or $\mu=0$ (the degenerate case being the ``hyperbola-like'' curve $T^{(1-C_2)}\mu^{-(1-C_1)}=0$), or they are power law curves that go through the origin at $T=0,\mu=0$ (degenerate cases given by $T=0$ or $\mu=0$), depending on the particular values of $C_1$ and $C_2$.  For us we have $C_1=C_2$ so that lines of constant $\frac{T}{\mu}$ are related, which matches the scaling argument above.

Note that the curve $\mu=0$ is a characteristic curve, simply setting $\mu_0=0$.  This means that the initial conditions provided by the curve $\mu=0$ are not sufficient to determine the functions $s$ and $n$ off of the curve $\mu=0$ (see for example \cite{Zachmanoglou}).  Further, no amount of additional perturbative information will yield a complete amount of information off a characteristic curve.  This is realized by the presence of the general function $\hat{g}$ in the general solutions (\ref{snslns}).  This basically comes down to saying something that we already knew: all relevant information differentiating models depends on $\frac{\mu^{\#}}{T}$ (again, $\#=1$ for AdS).  This is reflected in the geometry by saying that $\mu$ defines a scale, and $T$ defines a scale and all solutions related by rescaling $(\mu,T)\rightarrow (\lambda \mu, \lambda T)$  are related by the diffeomorphism $(r,t,x_i)\rightarrow \left(\lambda r,\lambda^{-1} t, \lambda^{-1} x_i\right)$ that leave the AdS asymptotics alone, but shift the relevant scales in the bulk.  What we have learned is that no amount of perturbative information around a solution can fix the thermodynamics.  This is good because there are many different holographic models with a $U(1)$ gauge invariance, all of which presumably have different thermal behavior.

Finally note that along a characteristic curve, the functions $\hat g$ and $\hat h$ are constant.  This allows us to quickly deduce that along a characteristic curve, $\sigma\rightarrow \sigma \hat{\lambda}$, $(T_0, \mu_0, s_0, n_0)$ $\rightarrow$ $(\hat{\lambda}^{1-C_1} T_0, \hat{\lambda}^{1-C_2} \mu_0, \hat{\lambda}^{C_1} s_0, \hat{\lambda}^{C_2} n_0)$.  For us $C_1=C_2= \frac{d}{d+1}$.  Calling $\hat{\lambda}=\lambda^{d+1}$ we find that the scaling becomes $(T_0, \mu_0, s_0, n_0)\rightarrow (\lambda T_0, \lambda \mu_0, \lambda^d s_0, \lambda^d n_0)$.  This is just the scaling in AdS$_{d+2}$ noted above (the scaling of ${\mc E}$ follows from (\ref{generalErelation})).

To summarize, we believe that the above arguments are general enough to apply to a theory where the only two scales are the chemical potential $\mu$ and the temperature $T$, and where the theory has a conformally invariant UV fixed point.  Under these conditions, we expect (\ref{generalErelation}) to hold, and that the different models are parameterized by the functions (say) $f\left(\frac{\mu}{T}\right)$.  This information is beyond relation (\ref{generalErelation}), i.e. it is information off of the characteristic curves that it defines.

We end by mentioning some possible future directions:
\begin{enumerate}
\item It would be interesting to explore the space of models with different functions $f\left(\frac{\mu}{T}\right)$.
\item One could study field theories on a $d$ dimensional sphere, i.e. the black hole geometries with spherical horizon.  In these cases we expect phase transitions.
\item Other dimensions:  one could expect there to be different thermal behavior, possibly phase transitions, for general $d$, as discussed at the end of section 1.
\item One could calculate certain correlation functions in the dual theory, and see how they interpolate between non-relativistic and relativistic behavior.
\item One could also study the dyonically charged versions of these black holes.  There should be constant dilaton solutions where the magnetic and electric charge are equal (see \cite{Gubankova:2010ny} for a recent study).
\item One could also study theories that asymptote to the Lifshitz-like background and are charged under a different $U(1)$ field (essentially adding a scale in the Lifshitz background).  However, note the curious result in \cite{Pang:2009pd} where an analytic charged solution is found, but the potential diverges.  It would be interesting to see if there are solutions with a finite (chemical) potential.  Using similar scaling arguments, we expect that a relationship of the type (\ref{nicerelation}) can be written in such a case with $C_1=C_2=\frac{d}{d+z}$.  However, we would like to back this up with an explicit calculation.

\end{enumerate}

\section*{Acknowledgements}
We wish to thank Bob Holdom for programming tips in maple, and Ida G. Zadeh who was involved in early stages of this project.  This work was supported by NSERC Canada.

\end{document}